\documentclass[preprint,review,12pt]{elsarticle}
\usepackage{lineno}
\usepackage{graphicx}
\usepackage{amsmath}
\usepackage{amssymb}
\usepackage{setspace}
\usepackage{enumerate}
\usepackage{microtype}
\usepackage{xcolor}
\usepackage{hyperref}
\definecolor{deepgreen}{rgb}{0,0.5,0}
\hypersetup{
    colorlinks=true,
    citecolor={deepgreen}
}
\usepackage{color}
\usepackage{subcaption}
\usepackage{array}
\usepackage{url}

\usepackage[ruled,linesnumbered,vlined]{algorithm2e}

\usepackage{multirow}
\newcolumntype{C}[1]{>{\centering\let\newline\\\arraybackslash\hspace{0pt}}m{#1}}

\journal{arXiv}

\begin{document}

\begin{frontmatter}

\title{Toroidal density-equalizing map for genus-one surfaces}

\author[label1]{Shunyu Yao}
\ead{syyao@link.cuhk.edu.hk}

\author[label1]{Gary P. T. Choi\corref{cor1}}
\cortext[cor1]{Corresponding author.}
\ead{ptchoi@cuhk.edu.hk}

\address[label1]{Department of Mathematics, The Chinese University of Hong Kong, Hong Kong}

\begin{abstract}
Density-equalizing map is a shape deformation technique originally developed for cartogram creation and sociological data visualization on planar geographical maps. In recent years, there has been an increasing interest in developing density-equalizing mapping methods for surface and volumetric domains and applying them to various problems in geometry processing and imaging science. However, the existing surface density-equalizing mapping methods are only applicable to surfaces with relatively simple topologies but not surfaces with topological holes. In this work, we develop a novel algorithm for computing density-equalizing maps for toroidal surfaces. In particular, different shape deformation effects can be easily achieved by prescribing different population functions on the torus and performing diffusion-based deformations on a planar domain with periodic boundary conditions. Furthermore, the proposed toroidal density-equalizing mapping method naturally leads to an effective method for computing toroidal parameterizations of genus-one surfaces with controllable shape changes, with the toroidal area-preserving parameterization being a prime example. Experimental results are presented to demonstrate the effectiveness of our proposed methods.

\end{abstract}


\end{frontmatter}

\section{Introduction}
The density-equalizing map is a technique proposed by Gastner and Newman~\cite{gastner2004diffusion} for cartogram creation. It produces a shape deformation on a planar domain based on the physical principle of density diffusion. Over the past few decades, the technique has been widely used for the visualization and analysis of sociological data including global epidemics~\cite{colizza2006role}, species extinction~\cite{wake2008we}, climate warming~\cite{lenoir2020species}, democratization~\cite{gleditsch2006diffusion}, and social media usage~\cite{mislove2011understanding}. In recent years, there have been continuous efforts in the development and analysis of the density-equalizing map and its variants. For instance, Gastner et al.~\cite{gastner2018fast} developed a fast flow-based method for computing density-equalizing maps with improved efficiency. Several works have also studied the effectiveness of contiguous area cartograms~\cite{duncan2020task,fung2024effectiveness}. Another recent direction has been the development of density-equalizing mapping methods for more general domains. In~\cite{choi2018density}, Choi and Rycroft proposed a mapping method for simply connected open surfaces based on the density-equalizing map technique. Later, Lyu et~al.~\cite{lyu2024bijective} developed an extension of the density-equalizing mapping method for multiply connected open surfaces. Some other recent works have also developed density-equalizing mapping methods for genus-0 closed surfaces~\cite{li2018diffusion,lyu2024spherical,lyu2024ellipsoidal} and volumetric domains~\cite{li2019visualization,choi2021volumetric}. Many of the above-mentioned surface and volumetric density-equalizing mapping methods have been applied to medical imaging~\cite{choi2020area}, volumetric data visualization~\cite{choi2021volumetric} and shape analysis~\cite{shaqfa2024disk,choi2024hemispheroidal}. 

As pointed out in~\cite{choi2018density}, the computation of density-equalizing maps is closely related to the problem of surface parameterization, which refers to the process of mapping a complicated surface onto a simple domain with some prescribed mapping criteria. Over the past several decades, numerous surface parameterization methods have been developed~\cite{floater2005surface,sheffer2007mesh}. In particular, most prior works have only focused on simply connected open surfaces (topological disk) and genus-0 closed surfaces (topological sphere), while the surface parameterization of more topologically complex surfaces is much less studied. For genus-one surfaces, Steiner and Fischer~\cite{steiner2004planar} developed a method for the planar parameterization of genus-one closed surfaces. Also, Ray et al.~\cite{ray2006periodic} developed a periodic global parameterization method that can be applied to genus-one surfaces. Zeng et al.~\cite{gu2007conformal} developed a method for conformally mapping a genus-one surface onto a two-layered sphere. Lam et al.~\cite{lam2015landmark} developed a method for landmark-constrained genus-one surface mapping using Teichm\"uller extremal maps. More recently, Yueh et al.~\cite{yueh2020new} proposed a volume-preserving parameterization method for genus-one three-manifolds. 

\begin{figure}[t!]
    \centering
    \includegraphics[width=\linewidth]{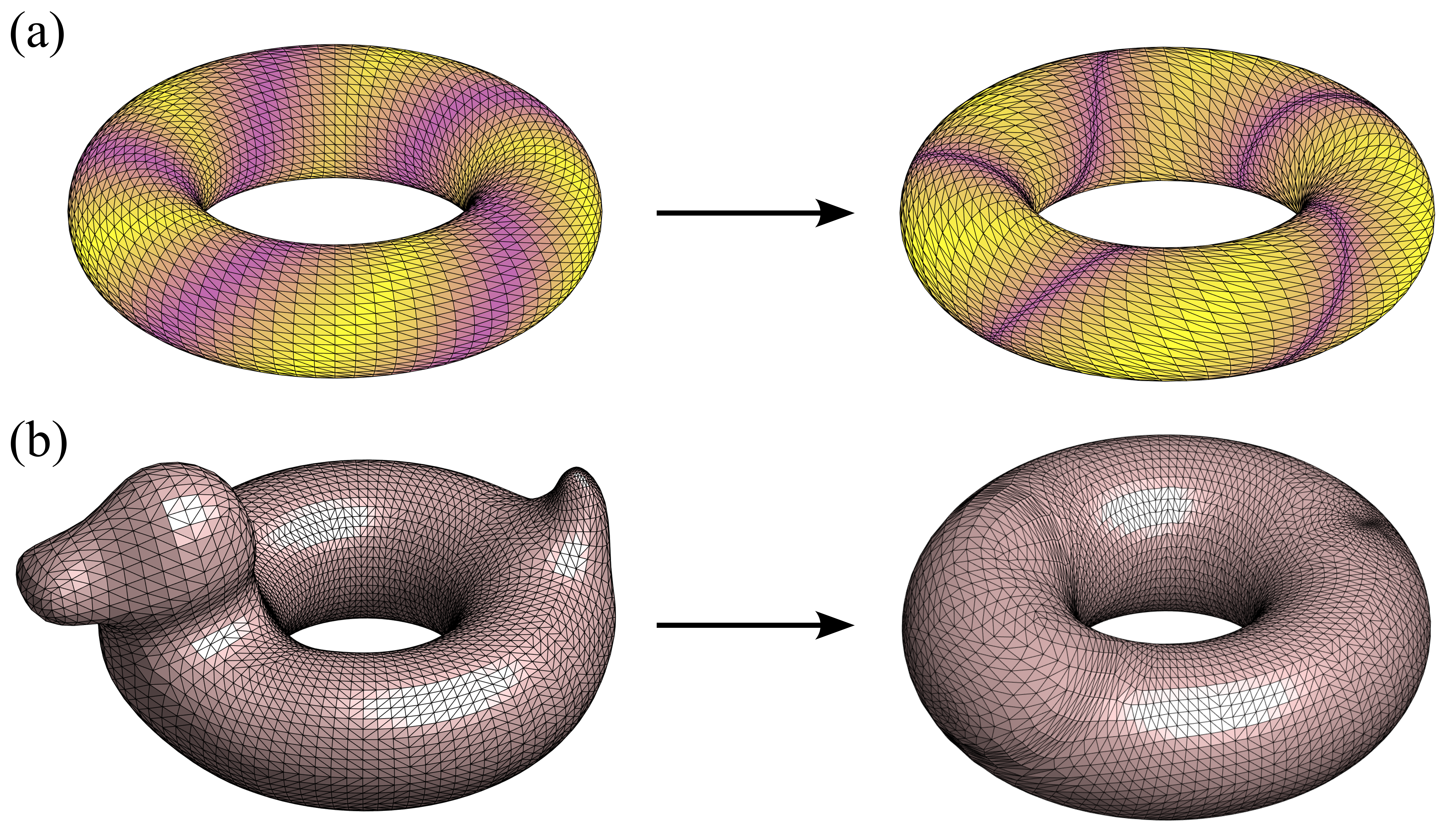}
    \caption{An illustration of the proposed toroidal density-equalizing map and toroidal parameterization methods. (a) Using the proposed toroidal density-equalizing map (TDEM) algorithm, we can achieve shape deformations on a toroidal surface with different regions enlarged or shrunk based on some given density information. (b) Using the proposed toroidal density-equalizing parameterization method, we can effectively map a genus-one surface onto a prescribed torus with controllable area changes.}
    \label{fig:illustration}
\end{figure}

While density-equalizing mapping techniques have been extensively studied and used for problems involving simple topologies, their high-genus analog has not been explored. In this work, we develop a novel method for computing density-equalizing maps for toroidal surfaces (abbreviated as TDEM) based on the principle of density diffusion (Fig.~\ref{fig:illustration}(a)). Using the proposed method, we can create different shape deformation effects on the given toroidal surface while preserving its overall geometry by simply changing the input density. We then further develop a toroidal density-equalizing parameterization method for general genus-one surfaces based on the proposed TDEM method, which allows us to easily control the local area changes and achieve toroidal area-preserving parameterizations (Fig.~\ref{fig:illustration}(b)).

The organization of this paper is as follows. In Section~\ref{sec:background}, we introduce the basic concepts of density diffusion and the formulation of density-equalizing maps. In Section~\ref{sec:tdem}, we present our proposed TDEM method for computing toroidal density-equalizing maps. In Section~\ref{sect:parameterization}, we describe our proposed toroidal parameterization method for general genus-one surfaces based on the proposed TDEM method. In Section~\ref{sec:experiment}, we present various numerical experiments to assess the performance of our proposed methods and demonstrate their effectiveness. We conclude our work and discuss possible future directions in Section~\ref{sec:conclusion}.

\section{Density diffusion and density-equalizing maps} \label{sec:background}
In this section, we describe the basic principles of density diffusion and density-equalizing maps. 

\begin{figure}[t]
    \centering
    \includegraphics[width=\linewidth]{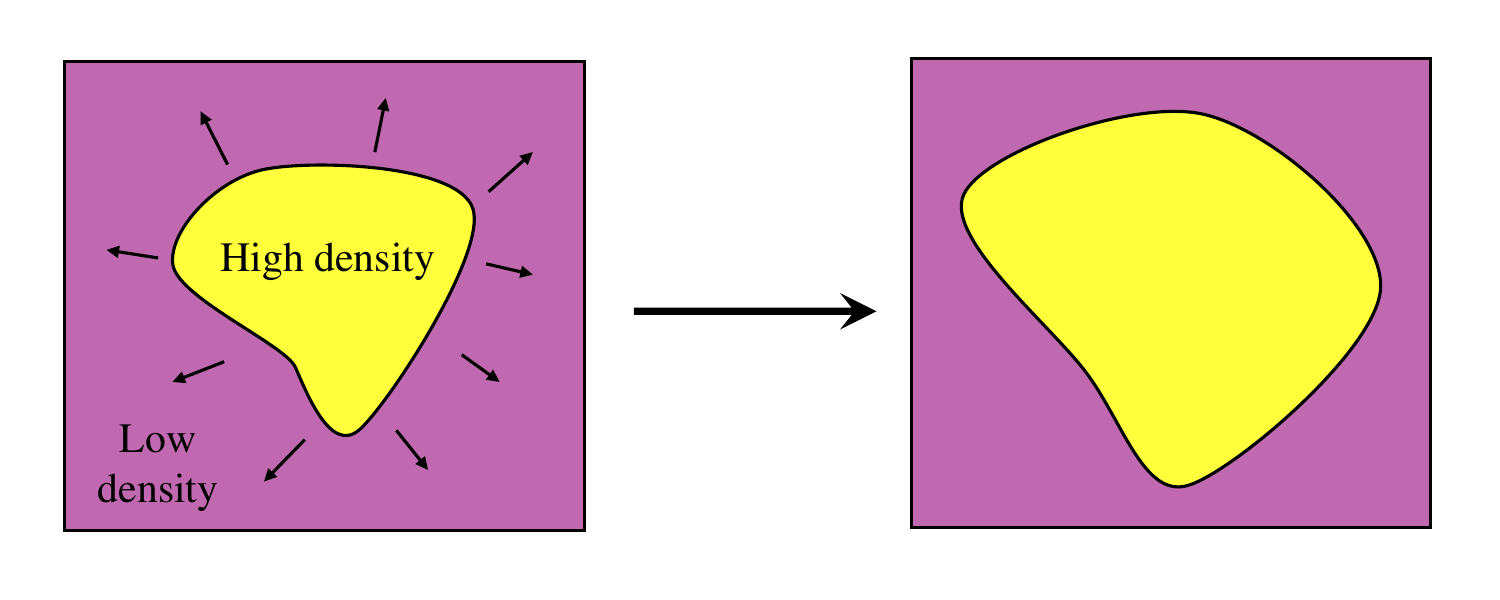}
    \caption{An illustration of density-equalizing maps. Given a planar domain $\mathcal{D}$ with some high-density regions (yellow) and some low-density regions (purple), the density-equalizing map will produce a shape deformation based on the density diffusion process. In particular, the high-density regions will expand and the low-density regions will shrink.}
    \label{fig:dem}
\end{figure}

The density-equalizing mapping method was originally developed by Gastner and Newman~\cite{gastner2004diffusion} as a diffusion-based cartogram creation technique. Given a planar domain $\mathcal{D}$ representing a geographical map and a positive ``population'' function defined on every part of the domain, where the ``population'' can be human population, income, GDP, or other given data. The density-equalizing map aims to deform $\mathcal{D}$ to equalize the density $\rho(\mathbf{x},t)$, which is defined as the population of position $\mathbf{x}$ divided by its area and will be changed as time $t$ flows. We follow the principle of density diffusion to equalize $\rho$ and have the advection equation
\begin{equation}\label{eq:advection}
    \frac{\partial \rho}{\partial t} + \nabla \cdot \mathbf{J} = 0,
\end{equation}
where $\mathbf{J} = \rho(\mathbf{x},t) \mathbf{v}(\mathbf{x},t)$ is the flux and $\mathbf{v}(\mathbf{x},t)$ is the velocity vector. The flux $\mathbf{J}$ can also be given by the Fick's law:
\begin{equation}\label{eq:fick's law}
    \mathbf{J} = -D \nabla \rho.
\end{equation}
In the above equation, the diffusion coefficient $D$ can be set to $1$ since we can always rescale the time properly.
Then, combining Eq.~\eqref{eq:advection} and Eq.~\eqref{eq:fick's law}, we have the diffusion equation
\begin{equation}\label{eq:diffusion}
    \frac{\partial \rho}{\partial t} = \Delta \rho
\end{equation}
and the velocity equation
\begin{equation}\label{eq:velocity}
    \mathbf{v} = - \frac{\nabla \rho}{\rho}.
\end{equation}
Since the velocity $\mathbf{v}$ is independent of the absolute scale of $\rho$, the update of the velocity field is stable.
Note that any tracer $\mathbf{x}(t)$ will move with the velocity $\mathbf{v}(\mathbf{x}(t),t)$.
We have the tracer displacement equation
\begin{equation}\label{eq:displacement}
    \mathbf{x}(t) = \mathbf{x}(0) + \int_0^t \mathbf{v}(\mathbf{x}(\tau),\tau) d \tau.
\end{equation}
After the diffusion, the density $\rho$ will be equal at every position in the deformed domain and hence $\nabla \rho \equiv 0$.
This means that $\lim_{t \rightarrow \infty} \mathbf{x}(t)$ will exist for any tracer $\mathbf{x}(t)$ and it will give the stable position of the corresponding part of the domain (see Fig.~\ref{fig:dem} for an illustration). To avoid infinite expansion of the domain, Gastner and Newman~\cite{gastner2004diffusion} put a large auxiliary rectangular region as the ``sea'' outside the domain $\mathcal{D}$ and set the density at the sea as the average of the density at $\mathcal{D}$. The density diffusion and mapping computation are performed on the entire region including both $\mathcal{D}$ and the auxiliary sea. This effectively constrains the diffusion so that the actual domain of interest can undergo shape deformations based on the prescribed density information without expanding infinitely.

While the density-equalizing mapping approach was originally developed for sociological data visualization, it was later extended and utilized as a method for surface mapping and parameterization~\cite{choi2018density}.

\section{Toroidal density-equalizing maps}\label{sec:tdem}

Consider a torus
\begin{equation}
    \mathcal{T} = \left\{(X,Y,Z) \in \mathbb{R}^3 \left| \left(\sqrt{X^2+Y^2} - R\right)^2 \right. + Z^2 = r^2 \right\},
\end{equation}
where $R$ and $r$ are the prescribed major and minor radii with $R > r > 0$. Let $\rho(X,Y,Z)$ be a density function defined on $\mathcal{T}$. Our goal is to compute a density-equalizing map $f: \mathcal{T} \to \mathcal{T}$ that produces a deformation on the toroidal domain based on some prescribed density. More specifically, let $\rho$ be a density function on the torus $\mathcal{T}$ defined as the ``population'' per unit area. Here, the ``population'' is some positive quantity (analogous to the human population in the original cartogram creation problem~\cite{gastner2004diffusion}) that we prescribe to create a desired shape deformation effect. In the discrete case, we represent the torus $\mathcal{T}$ as a triangulated surface $(\mathcal{V}, \mathcal{E}, \mathcal{F})$ where $\mathcal{V}$ is the set of vertices, $\mathcal{E}$ is the set of edges and $\mathcal{F}$ is the set of triangular faces. The density $\rho$ is discretized on $\mathcal{T}$ as 
\begin{equation}
    \rho(T) = \frac{P(T)} {\text{Area}(T)},
\end{equation}
where $T$ is a triangular face in $\mathcal{F}$ and $P(T)$ is the prescribed population function.

Intuitively, one may directly extend the computation in Section~\ref{sec:background} to the toroidal case by solving the diffusion equation on the torus and performing the vertex position update accordingly. However, the density gradient $\nabla \rho$ in $\mathbb{R}^3$ may lead to a change in the toroidal geometry. Specifically, under the vertex update, some vertices may no longer lie on the torus $\mathcal{T}$ and hence an extra projection step would be needed, which would lead to numerical errors and discrepancies between the density diffusion process and the shape deformations. Noticing that the torus is a parametric surface and can be easily represented by two angular coordinates, here we consider solving the density diffusion and shape deformation problems on a planar domain with periodic boundary conditions, thereby effectively mitigating the above-mentioned issues and simplifying the computational procedure.

We first define the toroidal projection $\phi: [0, 2\pi R)  \times [-\pi r, \pi r) \to \mathcal{T}$ as follows:
\begin{equation} \label{eq:toroidal}
\begin{split}
    \phi(u,v) &= \left(X(u,v),Y(u,v),Z(u,v)\right) \\
    &= \left(\left(R+r\cos \frac{v}{r}\right) \cos \frac{u}{R}, \left(R+r \cos \frac{v}{r}\right) \sin \frac{u}{R}, r \sin \frac{v}{r} \right).
\end{split}
\end{equation}
Note that $\phi$ gives a 1-1 correspondence between a rectangular domain with width $2\pi R$ and height $2\pi r$ and the torus $\mathcal{T}$. Also, it can be naturally extended for other values of $u,v$ outside the domain using the periodicity of the $\sin$ and $\cos$ functions. We remark that here we use $ [0, 2\pi R)  \times [-\pi r, \pi r)$ instead of the usual parameter domain with width $2\pi$ and height $2\pi$ so that the planar domain better resembles the geometry of the given torus. Also, we use $[-\pi r, \pi r)$ instead of $[0, 2\pi r)$ so that the top and bottom boundaries are mapped to the innermost loop of the torus. We remark that the inverse toroidal projection $\phi^{-1}: \mathcal{T} \to  [0, 2\pi R)  \times [-\pi r, \pi r)$ can also be explicitly expressed, with $\phi^{-1}(X,Y,Z) = \left(u(X,Y,Z),v(X,Y,Z)\right)$ where
\begin{equation}
    u(X,Y,Z) = \left\{ \begin{array}{ll}
        \displaystyle R \sin^{-1}\left(\frac{Y}{\sqrt{X^2+Y^2}}\right) & \ \text{ if } X \geq 0 \text{ and } Y \geq 0,\\
        \displaystyle  R \left(2\pi + \sin^{-1}\left(\frac{Y}{\sqrt{X^2+Y^2}}\right)\right) & \ \text{ if } X \geq 0 \text{ and } Y < 0,\\
        \displaystyle  R \left(\pi - \sin^{-1}\left(\frac{Y}{\sqrt{X^2+Y^2}}\right)\right) & \ \text{ otherwise,}\\
    \end{array}\right.
\end{equation}
with the range of the principal values of $\sin^{-1}$ being $[-\pi/2, \pi/2]$, and $v$ can be expressed in a similar manner:
\begin{equation}
    v(X,Y,Z) = \left\{ \begin{array}{ll}
        \displaystyle  r \sin^{-1}\left(\frac{Z}{r}\right) & \ \text{ if } X^2+Y^2 \geq R^2,\\
        \displaystyle  r \left(\pi-\sin^{-1}\left(\frac{Z}{r}\right)\right) & \ \text{ if } X^2+Y^2 < R^2  \text{ and } Z > 0,\\
        \displaystyle  -r \left( \pi+\sin^{-1}\left(\frac{Z}{r}\right) \right) & \ \text{ otherwise.}\\
    \end{array}\right.
\end{equation}

Now, we consider solving for the toroidal mapping $f$ via the planar domain instead of on the toroidal surface $\mathcal{T}$. Denote the initial position of all vertices on the plane as $\{\mathbf{x}_0(v_i)\}_i$. We first note that the toroidal projection $\phi$ is not an isometric mapping. Therefore, to ensure that the density diffusion on the planar domain is consistent with the density diffusion on the torus, we need to work with a modified density function that takes the difference between the planar domain and the torus into consideration. To achieve this, we define the modified density $\tilde{\rho}$ as
\begin{equation}
    \tilde{\rho}(T) = \frac{P(T)} {\text{Area}(\phi(T))},
\end{equation}
where $T = [\mathbf{x}_{0}(v_i),\mathbf{x}_{0}(v_j),\mathbf{x}_{0}(v_k)]$ is a triangle in the planar domain. Here, note that the area factor is computed based on the area of the mapped triangle $\phi(T)$ on the torus instead of the triangle $T$ on the plane. This ensures that while the density diffusion process and the associated vertex updates are performed on the plane subsequently, the resulting map will satisfy the desired shape deformation effect on the torus $\mathcal{T}$.

Now, we focus on performing the density diffusion and mapping processes on the plane based on the modified density $\tilde{\rho}$ using an iterative scheme. Analogous to Eq.~\eqref{eq:diffusion}, here we solve the diffusion equation 
\begin{equation}\label{eq:diffusion_modified}
        \frac{\partial \tilde{\rho}}{\partial t} = \Delta \tilde{\rho}.
\end{equation}
In the discrete case, the Laplacian is discretized as 
\begin{equation}
    \Delta = -A^{-1}L
\end{equation}
where $A$ is a $|\mathcal{V}|\times |\mathcal{V}|$ diagonal matrix (called the lumped area matrix) with its diagonal entries given by
\begin{equation}
    A_{ii} = \frac{1}{3} \sum_{T \in \mathcal{N}^{\mathcal{F}}(v_i)} \text{Area}(T),
\end{equation}
where $\mathcal{N}^{\mathcal{F}}(v_i)$ is the 1-ring face neighborhood of the vertex $v_i$, and $L$ is a $|\mathcal{V}|\times |\mathcal{V}|$ sparse matrix (called the cotangent Laplacian~\cite{pinkall1993computing}) with
\begin{equation}
L_{ij} = \left\{\begin{array}{ll}
    \displaystyle -\frac{1}{2}(\cot \alpha_{ij} + \cot \beta_{ij}) & \text{ if } i \neq j \text{ and } j \in \mathcal{N}^{\mathcal{V}}(v_i),\\
    \displaystyle  -\sum_{k=1}^{|\mathcal{V}|} L_{ik} & \text{ if } i = j,\\
    0 & \text{ otherwise,}
\end{array}\right.
\end{equation}
where $\mathcal{N}^{\mathcal{V}}(v_i)$ is the 1-ring vertex neighborhood of the vertex $v_i$, and $\alpha_{ij}, \beta_{ij}$ are two angles opposite to the the edge $[v_i, v_j] \in \mathcal{E}$ in the planar domain. We can then solve the diffusion equation in Eq.~\eqref{eq:diffusion_modified} using the semi-discrete backward Euler method iteratively:
\begin{equation}\label{eq:diffusion_modified_discrete}
    \frac{\tilde{\rho}_{n+1}-\tilde{\rho}_n}{\delta t} = -A^{-1} L \tilde{\rho}_{n+1} \Longleftrightarrow \tilde{\rho}_{n+1} = (A + L \delta t) ^{-1} A \tilde{\rho}_n,
\end{equation}
where $n$ is the iteration number and $\delta t$ is a prescribed step size.  

Recall that in the original density-equalizing mapping approach~\cite{gastner2004diffusion}, an auxiliary ``sea'' is set surrounding the actual planar domain to avoid infinite expansion of the domain and allow it to change freely in shape. Now, note that in our problem formulation for toroidal density-equalizing maps, the density diffusion process occurs on the plane, and the planar deformation result will be mapped back to the toroidal surface $\mathcal{T}$ via the toroidal projection $\phi$. Also, one can easily see that the top and bottom boundaries of the initial planar domain will be mapped to the same curve on $\mathcal{T}$ under $\phi$, as we have
\begin{equation}
\begin{split}
    &\phi(u,v+2\pi r) \\
    &= \left(\left(R+r\cos \frac{v+2\pi r}{r}\right) \cos \frac{u}{R}, \left(R+r \cos \frac{v+2\pi r}{r}\right) \sin \frac{u}{R}, r \sin \frac{v+2\pi r}{r} \right) \\
    &= \phi(u,v).
\end{split}
\end{equation}
Similarly, the left and right boundaries of the initial domain will be mapped to the same curve on $\mathcal{T}$ under $\phi$ as we have
\begin{equation}
\begin{split}
    &\phi(u+2\pi R,v) \\
    &= \left(\left(R+r\cos \frac{v}{r}\right) \cos \frac{u+2\pi R}{R}, \left(R+r \cos \frac{v}{r}\right) \sin \frac{u+2\pi R}{R}, r \sin \frac{v}{r} \right) \\
    &= \phi(u,v).
\end{split}
\end{equation}
Therefore, while the planar domain should change its overall shape under the density diffusion and shape deformation process, it is desired that the consistency between the boundaries when mapped back to the torus under toroidal projection will be preserved throughout the process. To achieve this, instead of introducing an auxiliary ``sea'' as in~\cite{gastner2004diffusion}, here we only need to consider the actual planar domain and modify the boundary constraints when solving the diffusion equation and performing the vertex update. 

\begin{figure}[t]
    \centering
    \includegraphics[width=\linewidth]{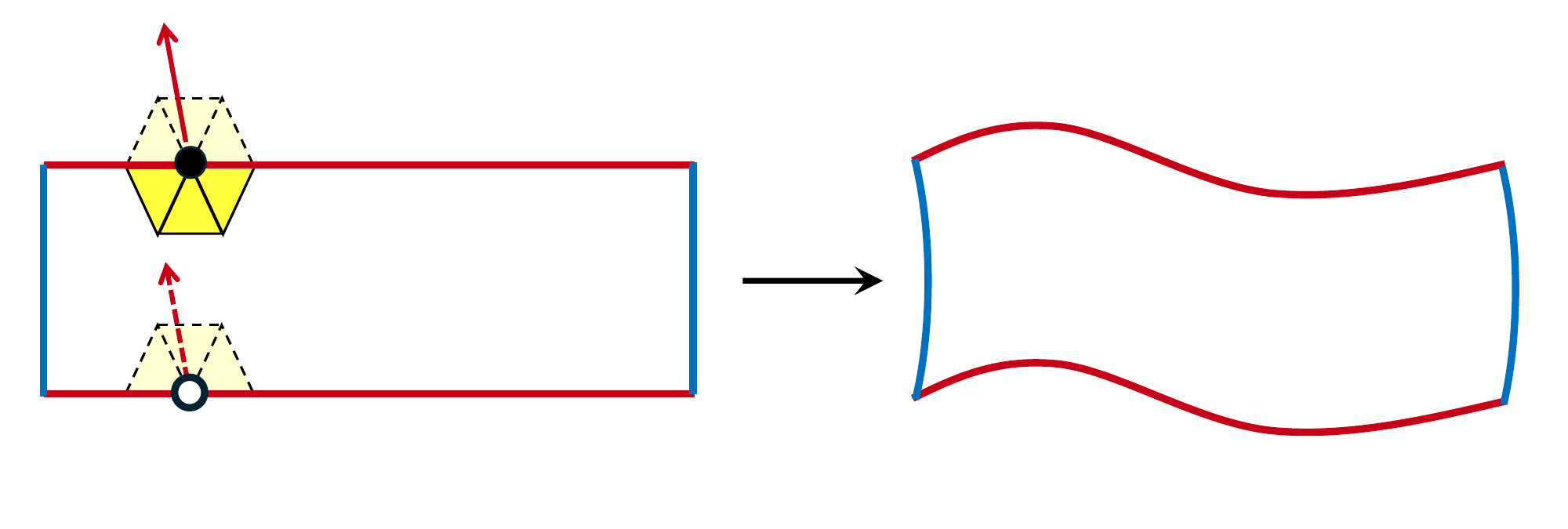}
    \caption{An illustration of the periodic boundary constraints in the planar density diffusion and mapping process. Throughout the planar density diffusion and mapping process, periodic boundary constraints need to be enforced for both the top and bottom boundaries (red) and both the left and right boundaries (blue). Specifically, for every pair of corresponding boundary vertices (the black dot and the white dot), a virtual copy of the neighboring nodes and triangles will be considered in the density diffusion process so that the density information can be exchanged along the boundaries. The boundaries are then updated consistently so that they only differ by a translation, which ensures that they will be mapped back to a consistent location on the torus under the toroidal projection.}
    \label{fig:periodic}
\end{figure}

Specifically, when solving the diffusion equation in Eq.~\eqref{eq:diffusion_modified_discrete}, we identify the opposite boundaries of the domain to ensure that the density information can be exchanged along the boundaries. This involves considering a virtual copy of the neighboring nodes and triangles for every pair of corresponding boundary vertices (see Fig.~\ref{fig:periodic} for an illustration) in the construction of the mass matrix $A$ and the cotangent Laplacian $L$. In practice, after computing the entries in a row in $A$ and $L$ associated with a vertex at the left or bottom boundary, we modify $A$ and $L$ by further adding the entries to the row associated with its corresponding vertex at the right or top boundary. We can then solve Eq.~\eqref{eq:diffusion_modified_discrete} to get the updated density $\tilde{\rho}_{n+1}$. We then copy the values of $\tilde{\rho}_{n+1}$ at the right and top boundary vertices to their counterparts at the left and bottom boundaries.

After getting the updated density $\tilde{\rho}_{n+1}$, we can compute the face-based density gradient for every face $T \in \mathcal{F}$ (see~\cite{choi2018density} for details):
\begin{equation}
\begin{split}
    &(\nabla \tilde{\rho}_{n+1})^{\mathcal{F}}(T) \\
    =& \frac{1}{2\text{Area}(T)} (\mathbf{n}(T) \times ((\tilde{\rho}_{n+1})_i (\mathbf{x}_{n}(v_k)-\mathbf{x}_{n}(v_j)) + (\tilde{\rho}_{n+1})_j (\mathbf{x}_{n}(v_i)-\mathbf{x}_{n}(v_k))  \\
    & + (\tilde{\rho}_{n+1})_k (\mathbf{x}_{n}(v_j)-\mathbf{x}_{n}(v_i)) )),
    \end{split}
\end{equation}
where $T = [\mathbf{x}_{n}(v_i),\mathbf{x}_{n}(v_j),\mathbf{x}_{n}(v_k)]$ is a triangular face, and $(\tilde{\rho}_{n+1})_i$, $(\tilde{\rho}_{n+1})_j$, $(\tilde{\rho}_{n+1})_k$ are the updated density values at the three vertices, and $\mathbf{n}(T)$ is the unit outward normal of the triangle $T$. Then, we can get the vertex-based density gradient from the face-based density gradient $(\nabla \tilde{\rho}_{n+1})^{\mathcal{F}}$ via a simple conversion:
\begin{equation}
    \nabla \tilde{\rho}_{n+1} = M (\nabla \tilde{\rho}_{n+1})^{\mathcal{F}},
\end{equation}
where $M$ is a $|\mathcal{V}| \times |\mathcal{F}|$ conversion matrix with
\begin{equation}\displaystyle
    M_{ij} =  \left\{\begin{array}{ll}
\displaystyle  \frac{\text{Area}(T_j)}{\sum_{T \in \mathcal{N}^{\mathcal{F}}(v_i)}\text{Area}(T)} & \text{ if $T_j$ is incident to $v_i$},\\
0 & \text{ otherwise.}
\end{array}\right.
\end{equation}
Analogous to the construction of $A$ and $L$, here we also encode the periodicity of the domain in the construction of $M$ by modifying the rows of it corresponding to the boundary vertices.

Now, from the vertex-based density gradient $\nabla \tilde{\rho}_{n+1}$, we can get the velocity field:
\begin{equation}\label{eq:velocity_torus}
    \mathbf{v} = -\frac{\nabla \tilde{\rho}_{n+1}}{\tilde{\rho}_{n+1}}.
\end{equation}
We then use the above velocity field to update the vertex positions on the plane:
\begin{equation} \label{eq:vertex_update_torus}
    \mathbf{x}_{n+1} = \mathbf{x}_{n} + \mathbf{v} \delta t,
\end{equation}
where $\mathbf{x}_{n}$ is the current position of the vertices and $\mathbf{x}_{n+1}$ is the updated position. As discussed above, there is a consistency between opposite boundaries in the initial planar domain as they correspond to the same curves under the toroidal projection. Because of the periodicity enforced in the above-mentioned diffusion equation solver and density gradient computation, the periodicity of the two pairs of boundaries in terms of their vertex positions under the vertex position update in Eq.~\eqref{eq:vertex_update_torus} is automatically satisfied (see Fig.~\ref{fig:periodic} for an illustration). Specifically, suppose $(x_{\text{top}}, y_{\text{top}})$ and $(x_{\text{bottom}}, y_{\text{bottom}})$ are two vertices on the top and bottom boundaries of the planar domain such that their corresponding positions on the torus $\mathcal{T}$ are identical, i.e., $\phi(x_{\text{top}}, y_{\text{top}}) = \phi(x_{\text{bottom}}, y_{\text{bottom}})$. Under the vertex position update, their positions on the plane will only differ by a translation:   
\begin{equation} \label{eqt:lr}
\left\{
\begin{array}{cl}
     x_{\text{top}} & = \  x_{\text{bottom}}, \\
     y_{\text{top}} & = \ y_{\text{bottom}} + 2\pi r.
\end{array}\right.
\end{equation}
Similarly, suppose $(x_{\text{left}}, y_{\text{left}})$ and $(x_{\text{right}}, y_{\text{right}})$ are two vertices on the left and right boundaries of the planar domain such that their corresponding positions on $\mathcal{T}$ are identical. Under the vertex position update, we will have:
\begin{equation} \label{eqt:tb}
\left\{
\begin{array}{cl}
     x_{\text{right}} & = \  x_{\text{left}} + 2\pi R, \\
     y_{\text{right}} & = \ y_{\text{left}}.
\end{array}\right.
\end{equation}
In other words, while the planar domain can change its shape from a rectangle to some other planar shape, the periodicity between the top and bottom boundaries and between the left and right boundaries is always preserved.

After performing the vertex update above, we follow the idea in~\cite{lyu2024spherical} and include a few extra steps for correcting potential mesh overlaps and recoupling the density and the deformation at the end of each iteration. Specifically, to correct local mesh fold-overs caused by large deformations induced by extreme density gradients, the mesh overlap correction scheme in~\cite{lyu2024spherical} involves solving a sparse linear system that updates the vertex positions. In our case, by fixing the boundary points, we can ensure that the periodicity will remain preserved under the correction step. Also, the density-deformation recoupling scheme~\cite{lyu2024spherical} aims to reduce numerical errors and discrepancies between the shape deformation and the density flow by re-defining the density using the updated vertex positions for the next iteration. In our case, we follow the same procedure and re-define the density for every triangle $T = [\mathbf{x}_{n+1}(v_i), \mathbf{x}_{n+1}(v_j), \mathbf{x}_{n+1}(v_k)]$ as 
\begin{equation}
\tilde{\rho}_{n+1}(T) = \frac{P(T)}{\text{Area}([\phi(\mathbf{x}_{n+1}(v_i)), \phi(\mathbf{x}_{n+1}(v_j)), \phi(\mathbf{x}_{n+1}(v_k))])}
\end{equation}
using the given population $P$ and the latest mapping result $\mathbf{x}_{n+1}$ at the end of each iteration. We can then repeat the above process until the density error $E(\tilde{\rho}_n) = \frac{\text{std}(\tilde{\rho}_n)}{\text{mean}(\tilde{\rho}_n)}$ is less than a prescribed error threshold $\epsilon$ or the number of iterations $n$ reaches a prescribed threshold $n_{\text{max}}$.

Finally, we can map the planar deformation result onto the torus $\mathcal{T}$ using the toroidal projection $\phi$ in Eq.~\eqref{eq:toroidal}. The resulting displacements of all vertices on $\mathcal{T}$ relative to their initial position on $\mathcal{T}$ give the desired density-equalizing map $f: \mathcal{T} \to \mathcal{T}$. The proposed TDEM algorithm is summarized in Algorithm~\ref{alg:tdem}. 

\begin{algorithm}[H]
 \KwData{A toroidal surface $\mathcal{T}$ with major radius $R$ and minor radius $r$, a prescribed population function $P$ defined on every triangular face, the step size $\delta t$, the error threshold $\epsilon$, and the maximum number of iterations $n_{\text{max}}$.}
 \KwResult{A toroidal density-equalizing map $f:\mathcal{T} \to \mathcal{T}$.}
 Use the inverse toroidal projection $\phi^{-1}$ to map $\mathcal{T}$ onto a planar domain $[0, 2\pi R) \times [-\pi r, \pi r)$\;
 Calculate the modified density $\tilde{\rho}_0(T) = \frac{P(T)}{\text{Area}([\phi(\mathbf{x}_0(v_i)), \phi(\mathbf{x}_0(v_j)),\phi(\mathbf{x}_0(v_k)]))}$ for all triangles $T = [\mathbf{x}_0(v_i), \mathbf{x}_0(v_j),\mathbf{x}_0(v_k)]$ on the plane\;
 \While{$E(\tilde{\rho}_n) \geq \epsilon$ and $n \leq n_{\text{max}}$}{
  Solve the diffusion equation in Eq.~\eqref{eq:diffusion_modified_discrete} with the periodicity of the boundaries considered in the construction of $A$ and $L$\;
  Compute the velocity field $\mathbf{v}$ using Eq.~\eqref{eq:velocity_torus}\;
  Update the vertex positions $\mathbf{x}_{n+1}$ using Eq.~\eqref{eq:vertex_update_torus}\;
  Apply the mesh overlap correction scheme\;
  Re-define the density as $\tilde{\rho}_{n+1}(T) = \frac{P(T)}{\text{Area}([\phi(\mathbf{x}_{n+1}(v_i)), \phi(\mathbf{x}_{n+1}(v_j)), \phi(\mathbf{x}_{n+1}(v_k))])}$ using the latest mapping result\;
  Update $n$ by $n+1$\;
 }
 Denote the planar mapping result as $g: \mathcal{D} \to \mathbb{R}^2$\;
 Apply the toroidal projection $\phi$ to map the planar mapping result onto $\mathcal{T}$\;
 The desired toroidal map is given by $f = \phi \circ g \circ \phi^{-1}$\;
 \caption{Toroidal density-equalizing map (TDEM)}
 \label{alg:tdem}
\end{algorithm}

\section{Toroidal density-equalizing parameterization of genus-one surfaces}\label{sect:parameterization}
Recall that the surface parameterization problem aims to map a given surface onto a simple standardized domain with some given mapping criteria. As the above toroidal density-equalizing mapping method is capable of creating a shape deformation on the torus based on some prescribed population functions, it is natural to extend the TDEM method for the parameterization of genus-one surfaces onto a given torus. 

More specifically, given a genus-one surface $\mathcal{S}$, a prescribed population defined on it, and a target toroidal surface $\mathcal{T}$ with major radius $R$ and minor radius $r$, our goal is to compute a toroidal parameterization $f: \mathcal{S} \to \mathcal{T}$ that achieves the desired area change effects encoded in the prescribed population. 

\begin{figure}[t!]
    \centering
    \includegraphics[width=\linewidth]{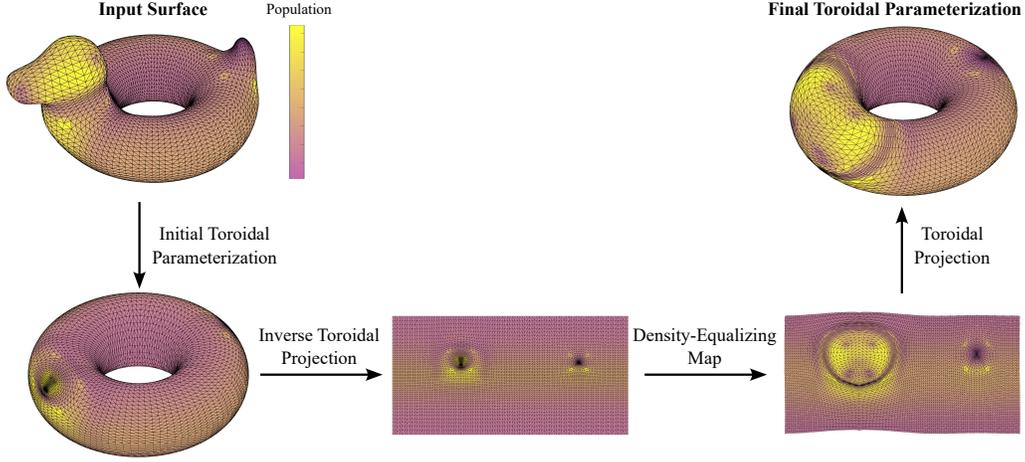}
    \caption{Outline of the proposed toroidal density-equalizing parameterization method. Given a genus-one surface with some prescribed population, we first compute an initial parameterization of it onto a torus. Then, we apply the inverse toroidal projection to map the torus onto the plane. We then apply our TDEM algorithm to obtain the deformed planar map with the periodic condition satisfied. Finally, we apply the toroidal projection to map the planar mapping result onto the torus, yielding the desired toroidal density-equalizing parameterization.}
    \label{fig:outline}
\end{figure}

Our proposed toroidal parameterization method is outlined in Fig.~\ref{fig:outline}. We first find an initial parameterization $h: \mathcal{S} \to \mathcal{T}$ that maps the input surface $\mathcal{S}$ onto the prescribed torus $\mathcal{T}$. The initial parameterization can be computed using any existing methods. In particular, a simple way is to slice $\mathcal{S}$ along two cut paths to get a simply connected open surface and map it onto a planar rectangular domain with periodic boundary conditions. In practice, we construct the initial toroidal parameterization by applying the rectangular conformal mapping method~\cite{meng2016tempo} with a modification of the boundary conditions as in~\cite{choi2021efficient}. Specifically, the method in~\cite{choi2021efficient} enforced the periodicity of the left and right boundaries, while here we enforce the periodicity for both the left and right boundaries and the top and bottom boundaries so that the mapping result gives a seamless initial toroidal parameterization.

After getting the initial toroidal parameterization, we can apply the inverse toroidal projection $\phi^{-1}$ to map the torus onto the plane. we can then apply the proposed TDEM method (Algorithm~\ref{alg:tdem}) to compute a toroidal density-equalizing map $g: \mathcal{T} \to \mathcal{T}$ via a planar mapping process with the periodic conditions satisfied. Here, the initial density $\tilde{\rho}$ for the TDEM method is given by $\tilde{\rho}(\hat{T}) = \frac{P(\hat{T})}{\text{Area}(\hat{T})}$, where $\hat{T}$ is a triangular face on the torus $\mathcal{T}$ associated with the triangular face $T$ on the input surface $\mathcal{S}$, i.e., $\hat{T} = h(T)$, and $P$ is the prescribed population function controlling the mapping effect.

Finally, the composition of the above mappings gives the desired toroidal parameterization $f = g \circ h$. The proposed toroidal density-equalizing parameterization method is summarized in Algorithm~\ref{alg:parameterization}.

\begin{algorithm}[H]
 \KwData{A genus-one surrface $\mathcal{S}$, the target toroidal surface $\mathcal{T}$, a prescribed population $P$, the step size $\delta t$, the error threshold $\epsilon$, and the maximum number of iterations $n_{\text{max}}$.}
 \KwResult{A toroidal parameterization $f:\mathcal{S} \to \mathcal{T}$.}
 Compute an initial parameterization $h: \mathcal{S} \to \mathcal{T}$\;
 Apply the TDEM algorithm with the prescribed population $P$, the step size $\delta t$, the error threshold $\epsilon$, and the maximum number of iterations $n_{\text{max}}$ to obtain a toroidal map $g: \mathcal{T} \to \mathcal{T}$\;
 The desired toroidal density-equalizing parameterization is given by $f =  g \circ h$\;
 \caption{Toroidal density-equalizing parameterization of genus-one surfaces}
 \label{alg:parameterization}
\end{algorithm}

As a particular case, the proposed toroidal parameterization algorithm can be used for producing toroidal area-preserving parameterization. Specifically, we follow the approach in~\cite{choi2018density} and define the input population as exactly the area of each triangular face of $\mathcal{S}$. Then, the initial density in the TDEM method will be given by $\tilde{\rho}(T) = \frac{\text{Area}(T)}{\text{Area}(h(T))}$, where $T$ is a triangular face on $\mathcal{S}$. By applying the TDEM algorithm to the torus based on $\tilde{\rho}$, the resulting map will give a constant final density of $\frac{\text{Area}(T)}{\text{Area}(g\circ h(T))} = \frac{\text{Area}(T)}{\text{Area}(f(T))}$. In other words, all triangles will be enlarged or shrunk precisely on the torus $\mathcal{T}$ according to their original size on $\mathcal{S}$, thereby yielding an area-preserving parameterization.

\section{Experimental results}\label{sec:experiment}
In this section, we present numerical experiments to evaluate the performance of our proposed toroidal density-equalizing map and toroidal parameterization methods. Various genus-one surface models from online mesh repositories~\cite{common,thingi10k,crane,edelsbrunner2001180web} are used in our experiments. The proposed algorithms are implemented in MATLAB R2021a. All experiments are performed on a Windows Computer with an Intel i9-12900 processor with 32GB RAM. In all our experiments, we set the time step size $\delta t = 0.1$, the error threshold $\epsilon = 10^{-3}$, and the maximum number of iterations $n_{\max} = 1000$.

\begin{figure}[t!]
    \centering
    \includegraphics[width=\linewidth]{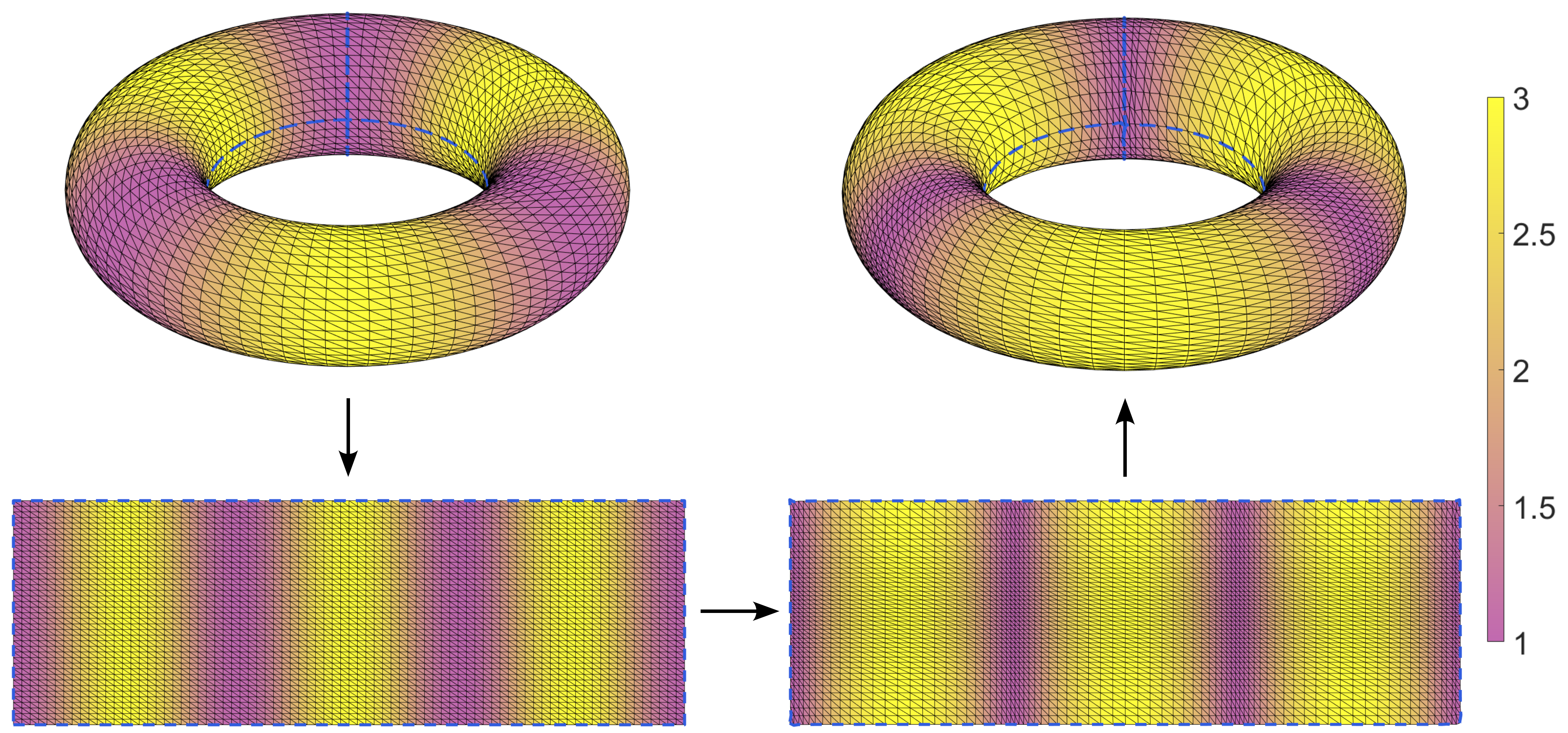}
    \caption{Mapping a torus with the prescribed population $P(T) = 2 - \cos(u)$, where $(u,v)$ is the face centroid of every triangle $T$ under the inverse toroidal projection $\phi^{-1}$. The top-left figure panel shows the original toroidal surface color-coded with the population. The bottom-left panel shows the planar representation of the original toroidal surface under $\phi^{-1}$. The bottom-right panel shows the planar mapping result. The top-right panel shows the final TDEM result. The blue dotted lines represent the cut paths. }
    \label{fig:example1}
\end{figure}

\subsection{Toroidal density-equalizing map}
We start by considering different density functions $\rho$ on a prescribed toroidal surface and applying the TDEM algorithm (Algorithm~\ref{alg:tdem}). 

In the first example (Fig.~\ref{fig:example1}), we consider a torus with major radius $R = 3$ and minor radius $r = 1$. We define the input population as $P(T) = 2 - \cos(u)$, where $(u,v)$ is the face centroid of every triangle $T$ under the inverse toroidal projection $\phi^{-1}$. This population function only involves variations along the toroidal direction but not the poloidal direction on the torus. Equivalently, there are only variations in the $x$-direction in the planar domain but not in the $y$-direction. We can easily see that the planar mapping does not produce any overall shape change, and the boundary vertices only slice along the boundaries. In the final TDEM result, we can also see that the shape deformation only involves enlargements and shrinkages along the toroidal direction.

\begin{figure}[t!]
    \centering
    \includegraphics[width=\linewidth]{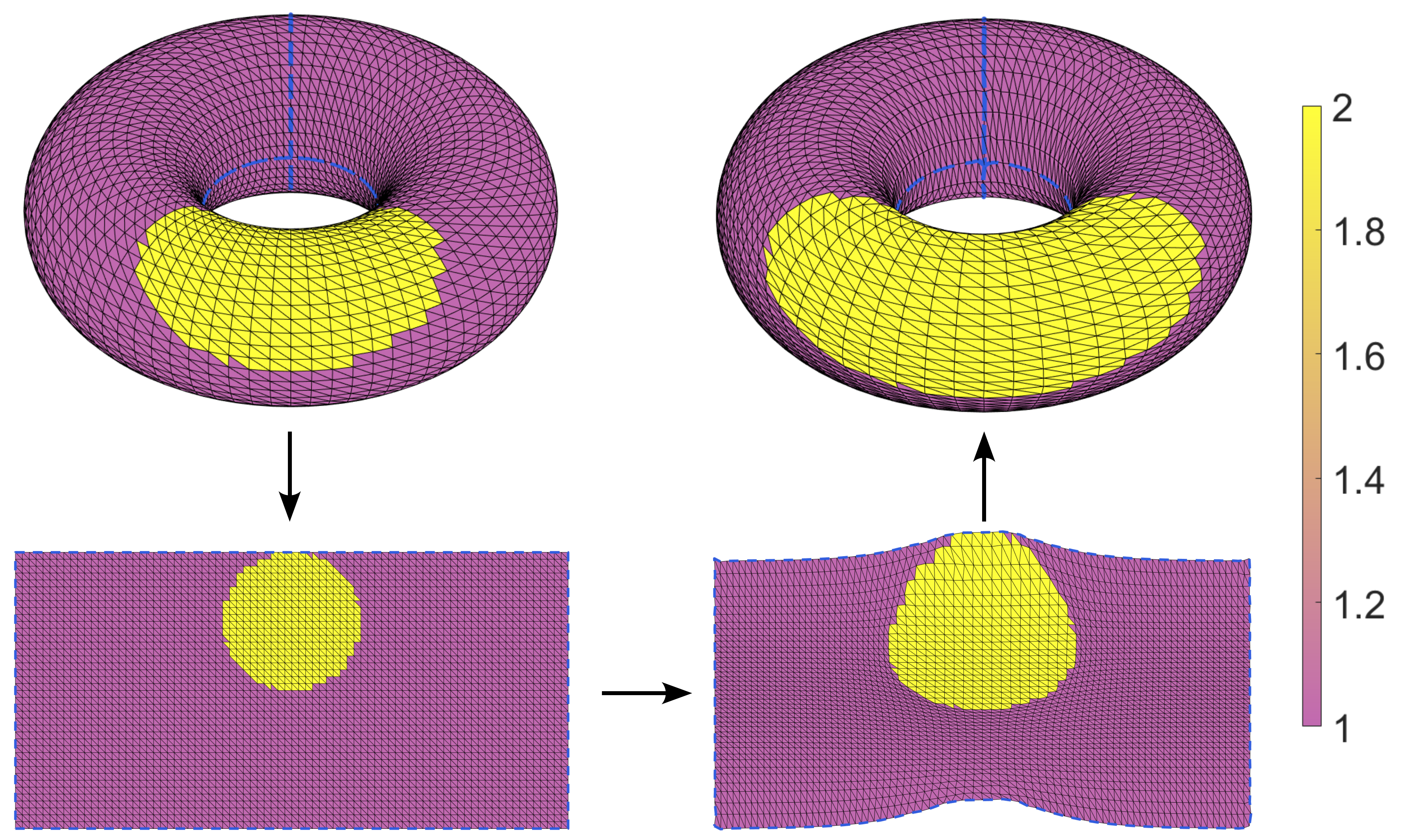}
    \caption{Mapping a torus with the prescribed population function involving a circular region with a higher population. The top-left figure panel shows the original toroidal surface color-coded with the population. The bottom-left panel shows the planar representation of the original toroidal surface under $\phi^{-1}$. The bottom-right panel shows the planar mapping result. The top-right panel shows the final TDEM result. The blue dotted lines represent the cut paths.}
    \label{fig:example2}
\end{figure}

In the second example (Fig.~\ref{fig:example2}), we consider a torus with $(R,r) = (2,1)$ and an input population $P$ where $P(T) = 2$ for the triangles corresponding to a ball centering at $(\pi R, \frac{\pi}{2}r)$ with the radius $\frac{r}{2}$ in the planar domain and $P(T) = 1$ otherwise. This population involves variations that produce boundary deformations in only one boundary pair of the planar domain. It can be observed in the planar mapping result that the top and bottom parts of the domain are significantly deformed in a periodic manner, while the overall shapes of the left and right boundaries are unchanged. In the final TDEM mapping result, we can see a clear deformation on the torus involving changes in both the toroidal and poloidal directions, where the high-density regions are expanded and the low-density regions are shrunk.

\begin{figure}[t!]
    \centering
    \includegraphics[width=\linewidth]{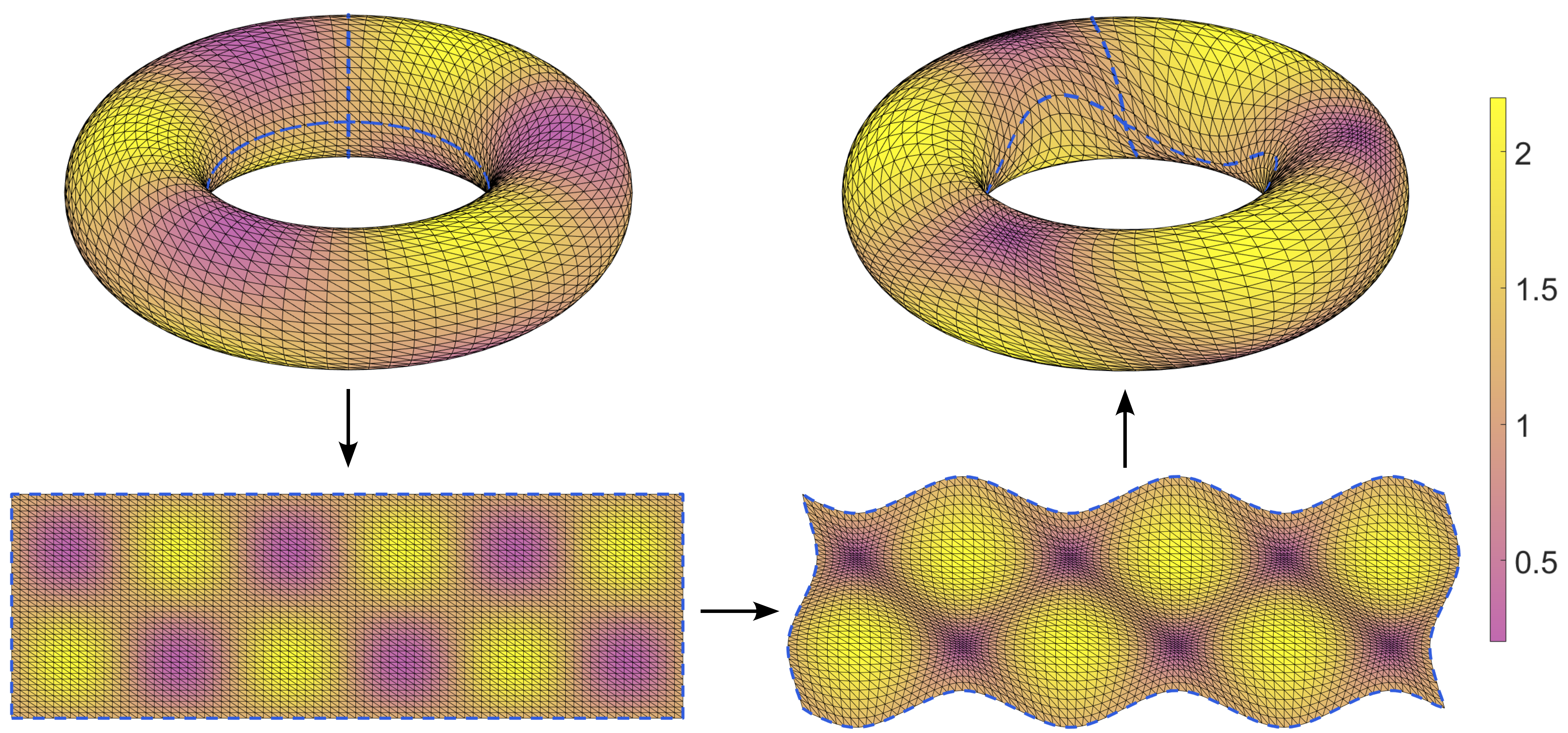}
    \caption{Mapping a torus with the prescribed population $P(T) = 1.2 - \sin(u) \sin(v)$, where $(u,v)$ is the face centroid of every triangle $T$ under the inverse toroidal projection $\phi^{-1}$. The top-left figure panel shows the original toroidal surface color-coded with the population. The bottom-left panel shows the planar representation of the original toroidal surface under $\phi^{-1}$. The bottom-right panel shows the planar mapping result. The top-right panel shows the final TDEM result. The blue dotted lines represent the cut paths.}
    \label{fig:example3}
\end{figure}

Next, we consider an example (Fig.~\ref{fig:example3}) for which the prescribed population is $1.2 - \sin(u) \sin(v)$ on a torus with $(R,r) = (3,1)$. Because of the choice of the population, the density involves variations that produce boundary deformations in both boundary pairs of the planar domain. As shown in the planar mapping result, both boundary pairs are deformed consistently to achieve density equalization. In the final TDEM result, it can also be observed that the vertices are deformed in multiple directions on the torus.

\begin{figure}[t!]
    \centering
    \includegraphics[width=\linewidth]{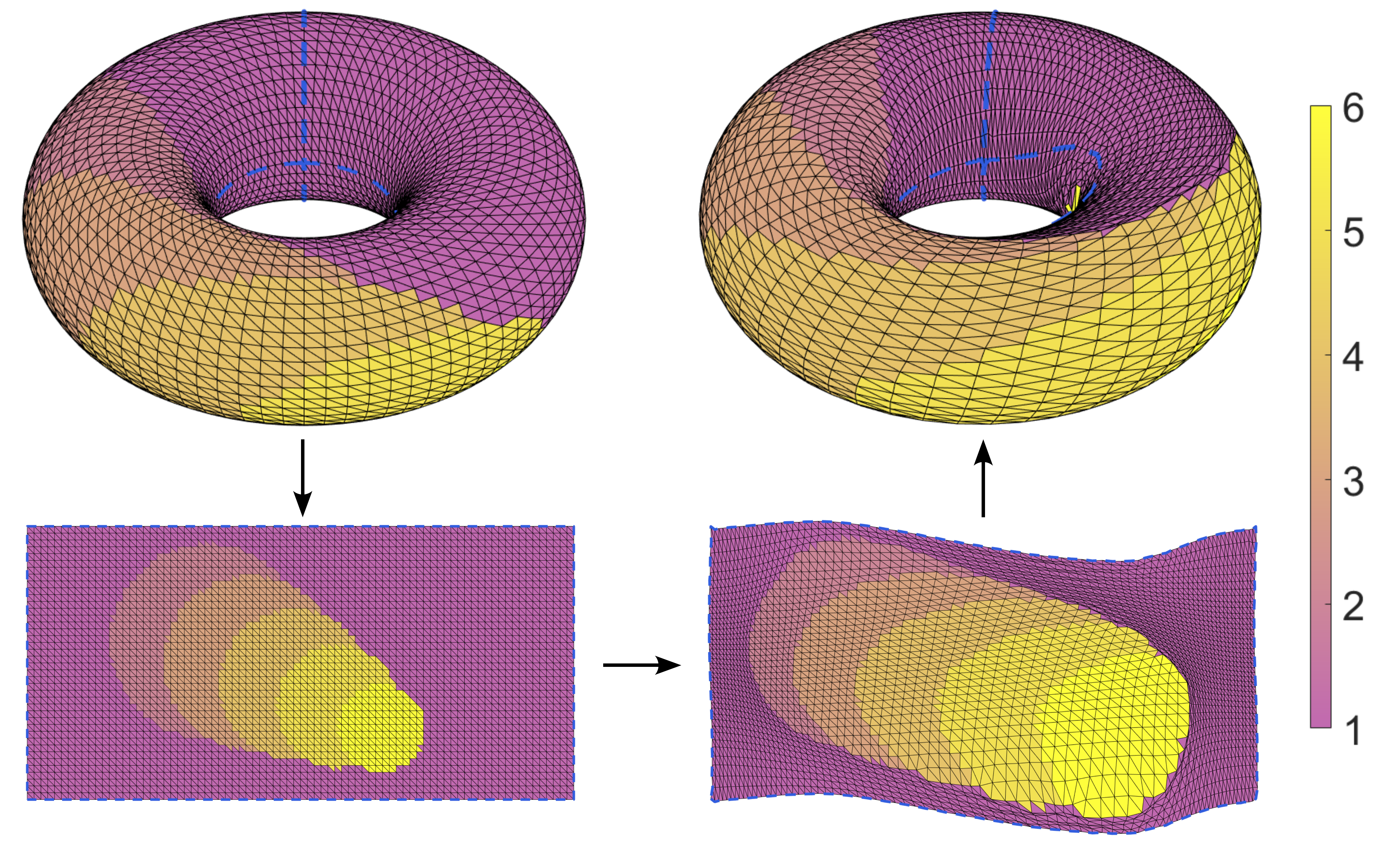}
    \caption{Mapping a torus with a more unsmooth input population. The top-left figure panel shows the original toroidal surface color-coded with the population. The bottom-left panel shows the planar representation of the original toroidal surface under $\phi^{-1}$. The bottom-right panel shows the planar mapping result. The top-right panel shows the final TDEM result. The blue dotted lines represent the cut paths.}
    \label{fig:example4}
\end{figure}

\begin{figure}[t!]
    \centering
    \includegraphics[width=\textwidth]{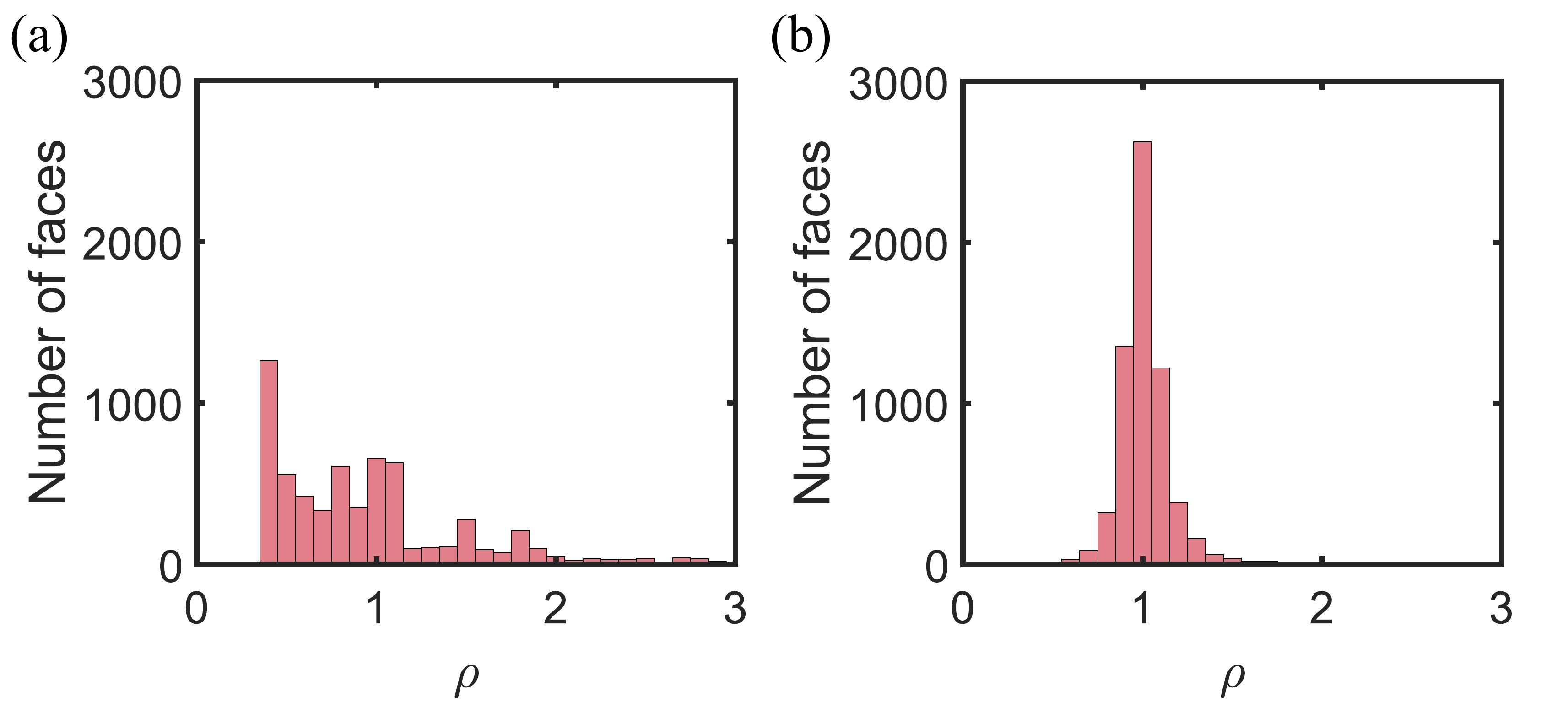}
    \caption{The initial and final density distributions in the example in Fig.~\ref{fig:example4}. (a) The initial density. (b) The final density.}
    \label{fig:distribution1}
\end{figure}

Note that the above examples involve some simple or smooth density distributions over the entire domain. It is natural to ask whether our TDEM method also works for more unsmooth densities. Here, we consider a more complicated example (Fig.~\ref{fig:example4}) with multiple overlapping circular regions with a higher population at the bottom right of the planar domain, while the populations at all other parts are uniform. In this case, it is expected that the bottom right part will expand because of the higher-density regions.  In the mapping result, we can see that indeed the diffusion process expands the bottom boundary and hence results in the top shrinking. Besides visualizing the mapping effect, we can also assess the mapping result by computing the density distributions before and after applying the TDEM method. The distribution of the density after the diffusion (Fig.~\ref{fig:distribution1}) is significantly central at $1$, which suggests that the density is effectively equalized.

\begin{figure}[t!]
    \centering
    \includegraphics[width=\linewidth]{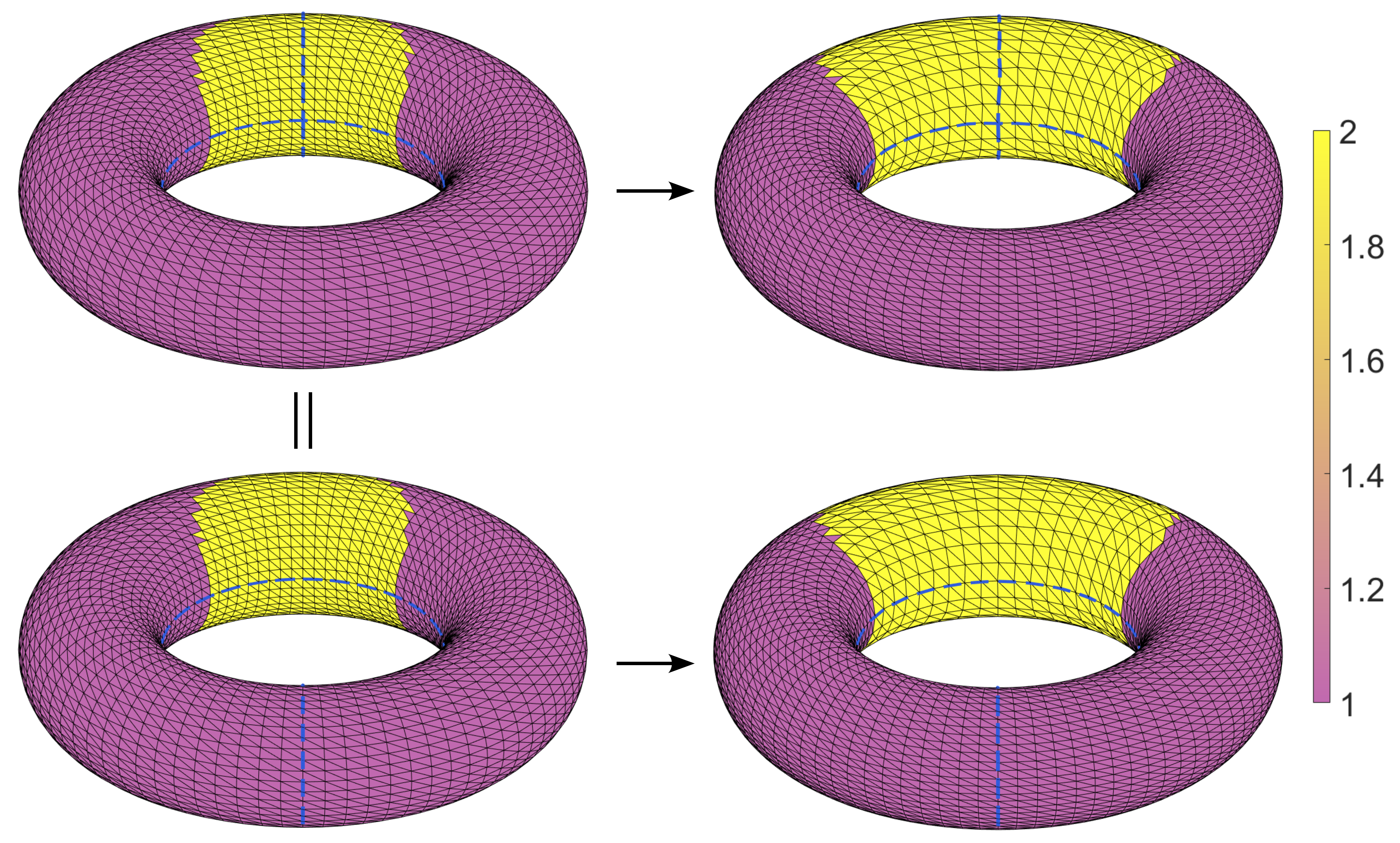}
    \caption{An example of mapping two toroidal surfaces with different cut paths. The left column shows two toroidal surfaces with an equivalent density distribution. The right column shows the final TDEM results. The blue dotted lines represent the cut paths.}
    \label{fig:example5}
\end{figure}

\begin{figure}[t!]
    \centering
    \includegraphics[width=\textwidth]{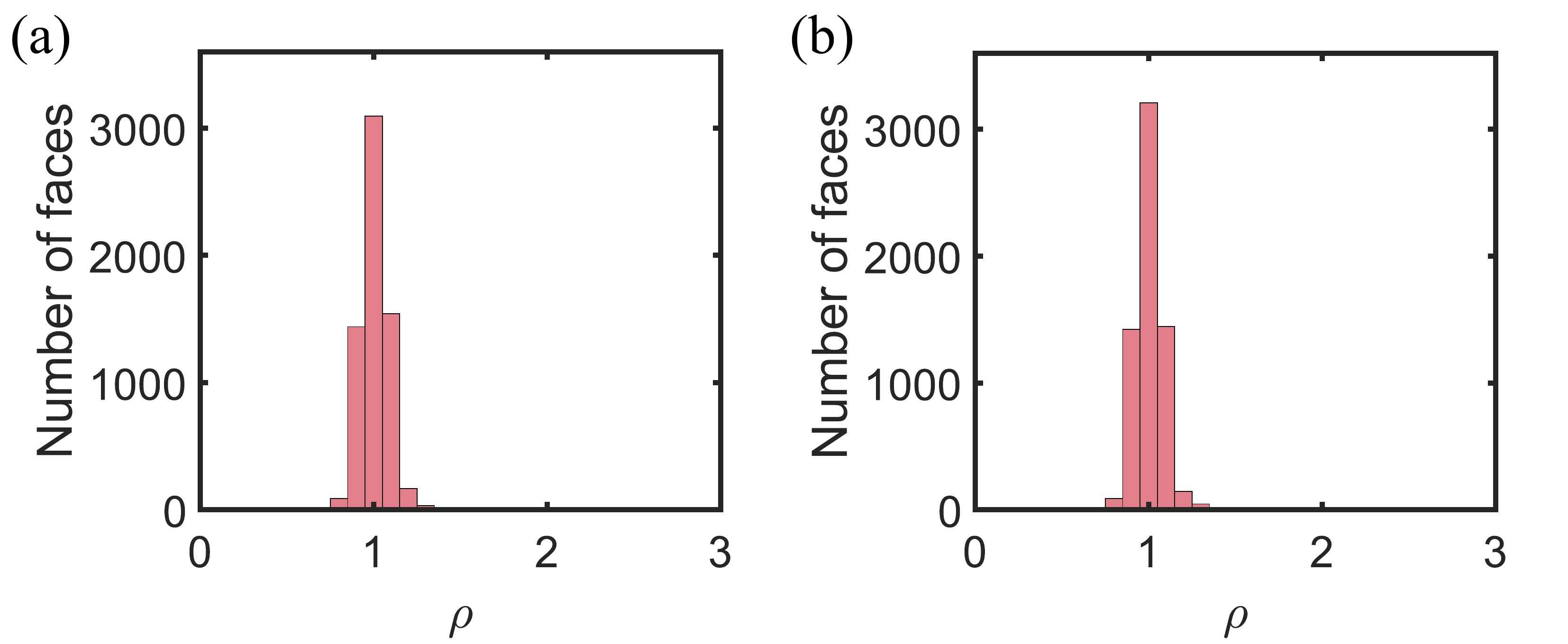}
    \caption{The final density distributions of the two TDEM mapping results with different cut paths in Fig.~\ref{fig:example5}. (a) The final density for the mapping example in which all the two cut paths pass through the center of the high-density region. (b) The final density for the mapping example in which only one of the cut paths passes through the center of the high-density region. It can be observed that the two distributions are highly consistent.}
    \label{fig:distribution2}
\end{figure}

Besides, it is necessary to check whether our TDEM method will return the same result of a certain density distribution on the torus with different cut paths. We consider a ball (Fig.~\ref{fig:example5}) on the torus with the population set to be $2$ inside and $1$ outside. The torus has radius $(R,r) = (3,1)$. In this example, we choose two different cut paths. The first choice is that all the cut paths will pass through the center of the ball while the second choice is that only one of them pass through the center. Because of the difference in the cut paths, the density distributions on the corresponding planar domains will be different. Nevertheless, under the TDEM mapping process, the periodicity of the planar domains will be enforced, and hence the two different planar mapping results will give two consistent toroidal density-equalizing maps. Specifically, it can be observed in Fig.~\ref{fig:distribution2} that the two TDEM mapping results are nearly identical.

For a more quantitative analysis, in Table~\ref{tab:tdem} we record the statistics of the mapping accuracy and computational efficiency of our TDEM method for all the above-mentioned examples. In particular, we consider the variance of the initial density and the variance of the final density, where both the initial and final densities are normalized so that the average density is 1. This allows us to make a fair comparison between the density variances to assess the density-equalizing effect of the mapping results. It can be observed in all examples that the variance of the final density is significantly smaller than that of the initial density, which suggests that our TDEM method achieves the density-equalizing effect very well.

\begin{table}[t!]
    \centering
    \begin{tabular}{c|c|c|c}
       Example  &  $\text{Var}(\bar{\rho}_{\text{initial}})$ & $\text{Var}(\bar{\rho}_{\text{final}})$ & Time (s)\\ \hline
       Fig.~\ref{fig:illustration}(a) & 0.2427 & $9.4623 \times 10^{-5}$ & 0.6963 \\ 
       Fig.~\ref{fig:example1} & 0.1959 & $2.1293\times 10^{-5}$ & 0.8316\\  
       Fig.~\ref{fig:example2} & 0.2180 & $2.1004\times 10^{-5}$ & 2.6590\\ 
       Fig.~\ref{fig:example3} & 0.2176 & $5.2278\times 10^{-5}$ & 0.7156\\ 
        Fig.~\ref{fig:example4} &  0.4569 & $5.4148\times 10^{-5}$ & 3.4841\\ 
        Fig.~\ref{fig:example5} (top) & $0.2085$ & $3.3402\times 10^{-6}$ & 5.7088\\ 
        Fig.~\ref{fig:example5} (bottom) & $0.2036$ & $4.2551\times 10^{-6}$ & 5.3904\\ 
    \end{tabular}
    \caption{The performance of our proposed toroidal density-equalizing map (TDEM) method. For each example, we record the variance of the initial density $\text{Var}(\bar{\rho}_{\text{initial}})$ where $\bar{\rho}_{\text{initial}} = \frac{{\rho}_{\text{initial}}}{\text{mean}(\rho_{\text{initial}})}$, the variance of the final density $\text{Var}(\bar{\rho}_{\text{final}})$ where $\bar{\rho}_{\text{final}} = \frac{{\rho}_{\text{final}}}{\text{mean}(\rho_{\text{final}})}$, and the time required for the TDEM algorithm. }
    \label{tab:tdem}
\end{table}

\begin{figure}[t!]
    \centering
    \includegraphics[width=\linewidth]{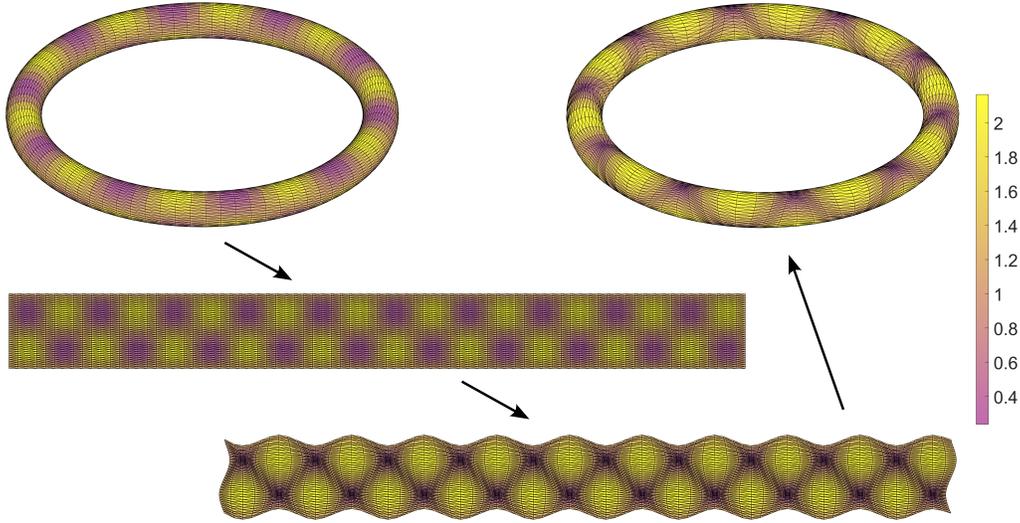}
    \caption{Mapping a torus with extreme radius parameters $(R,r) = (10, 1)$. The top-left figure panel shows the original toroidal surface color-coded with the population population $P(T) = 1.2 - \sin(u) \sin(v)$, where $(u,v)$ is the face centroid of every triangle $T$ under the inverse toroidal projection $\phi^{-1}$. The bottom-left panel shows the planar representation of the original toroidal surface under $\phi^{-1}$. The bottom-right panel shows the planar mapping result. The top-right panel shows the final TDEM result. }
    \label{fig:example_R10}
\end{figure}

One may also wonder whether our TDEM method can be applied to toroidal surfaces with different geometries. Here, we consider several tori with different radii and our TDEM method on them. For simplicity, we fix the minor radius as $r=1$ and vary the major radius $R = 2, 4, 6, 8, 10$. We then record the performance of our TDEM method on these toroidal surfaces with the input same population $P(T) = 1.2 - \sin(u) \sin(v)$ (see Fig.~\ref{fig:example_R10} for the example with $R = 10$). As shown in Table~\ref{tab:Rr}, our method can consistently produce accurate toroidal density-equalizing mapping results with the variance of the density significantly reduced. This suggests that our TDEM method is applicable to a wide range of toroidal surfaces with different geometries.

\begin{table}[t!]
    \centering
    \begin{tabular}{c|c|c|c|c}
       $R$  & $r$ & Var($\bar{\rho}_{\text{initial}}$) & Var($\bar{\rho}_{\text{final}}$) & Time (s)\\ \hline
        2 & 1 &  0.3045 & $1.2862 \times 10^{-4}$ & 0.7002 \\
        4 & 1 & 0.1910 & $4.1361 \times 10^{-5}$ & 0.7210 \\
        6 & 1 & 0.1699 & $4.0608 \times 10^{-5}$ & 0.7235 \\
        8 & 1 & 0.1583 & $5.2533 \times 10^{-5}$ & 0.7526 \\
        10 & 1 & 0.1483 & $7.9918 \times 10^{-5}$ & 1.0320 \\
    \end{tabular}
    \caption{The performance of our TDEM method for toroidal surfaces with different geometries. Here, we fix the minor radius $r = 1$ and consider various values of the major radius $R$. For each example, we record the variance of the initial density $\text{Var}(\bar{\rho}_{\text{initial}})$ where $\bar{\rho}_{\text{initial}} = \frac{{\rho}_{\text{initial}}}{\text{mean}(\rho_{\text{initial}})}$, the variance of the final density $\text{Var}(\bar{\rho}_{\text{final}})$ where $\bar{\rho}_{\text{final}} = \frac{{\rho}_{\text{final}}}{\text{mean}(\rho_{\text{final}})}$, and the time required for the TDEM algorithm. }
    \label{tab:Rr}
\end{table}

\subsection{Toroidal density-equalizing parameterization}
After demonstrating the effectiveness of our proposed TDEM method for computing toroidal density-equalizing maps, we test our proposed toroidal parameterization method for general genus-one surfaces.

\begin{figure}[t!]
    \centering
    \includegraphics[width=\linewidth]{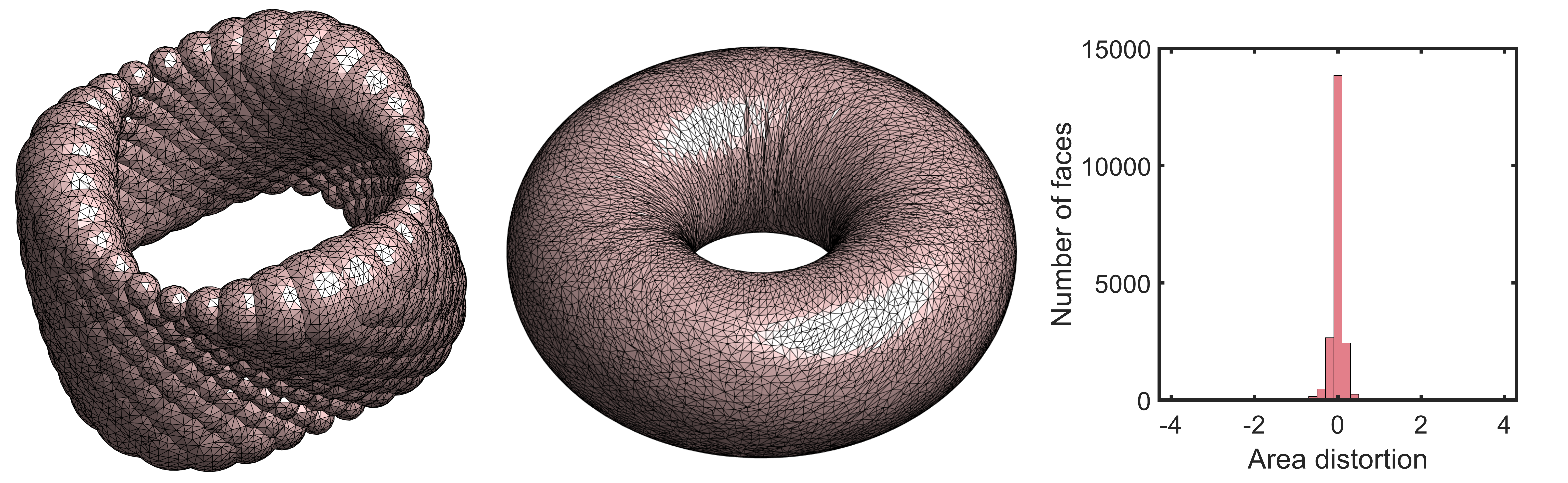}
    \caption{Toroidal area-preserving parameterization of the Bracelet model. Left to right: The input surface, the parameterization result by our algorithm, and the area distortion histogram.}
    \label{fig:parameterization_bracelet}
\end{figure}

\begin{figure}[t]
    \centering
    \includegraphics[width=\linewidth]{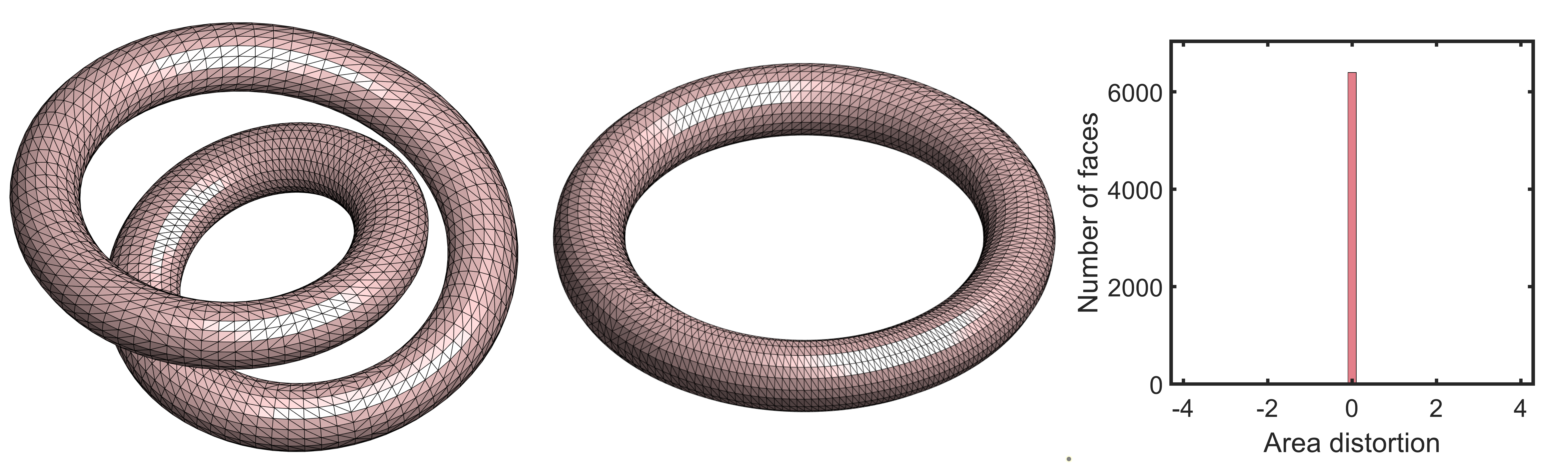}
    \caption{Toroidal area-preserving parameterization of the Wrapped Tube model. Left to right: The input surface, the parameterization result by our algorithm, and the area distortion histogram.}
    \label{fig:parameterization_2_1}
\end{figure}

\begin{figure}[t]
    \centering
    \includegraphics[width=\linewidth]{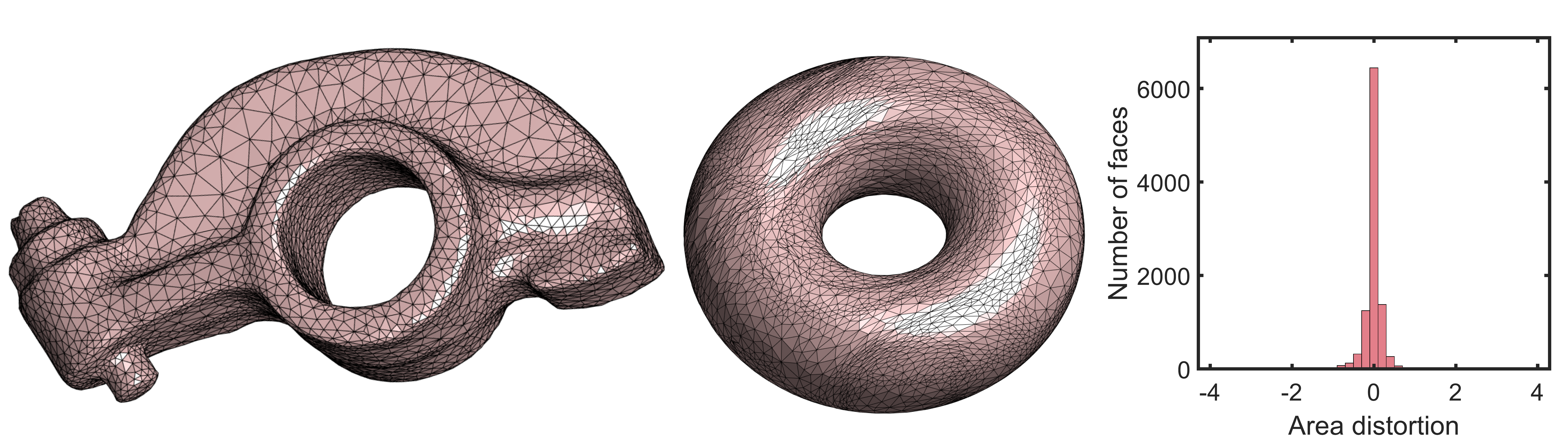}
    \caption{Toroidal area-preserving parameterization of the Rocker Arm model. Left to right: The input surface, the parameterization result by our algorithm, and the area distortion histogram.}
    \label{fig:parameterization_rocker_arm}
\end{figure}

\begin{figure}[t]
    \centering
    \includegraphics[width=\linewidth]{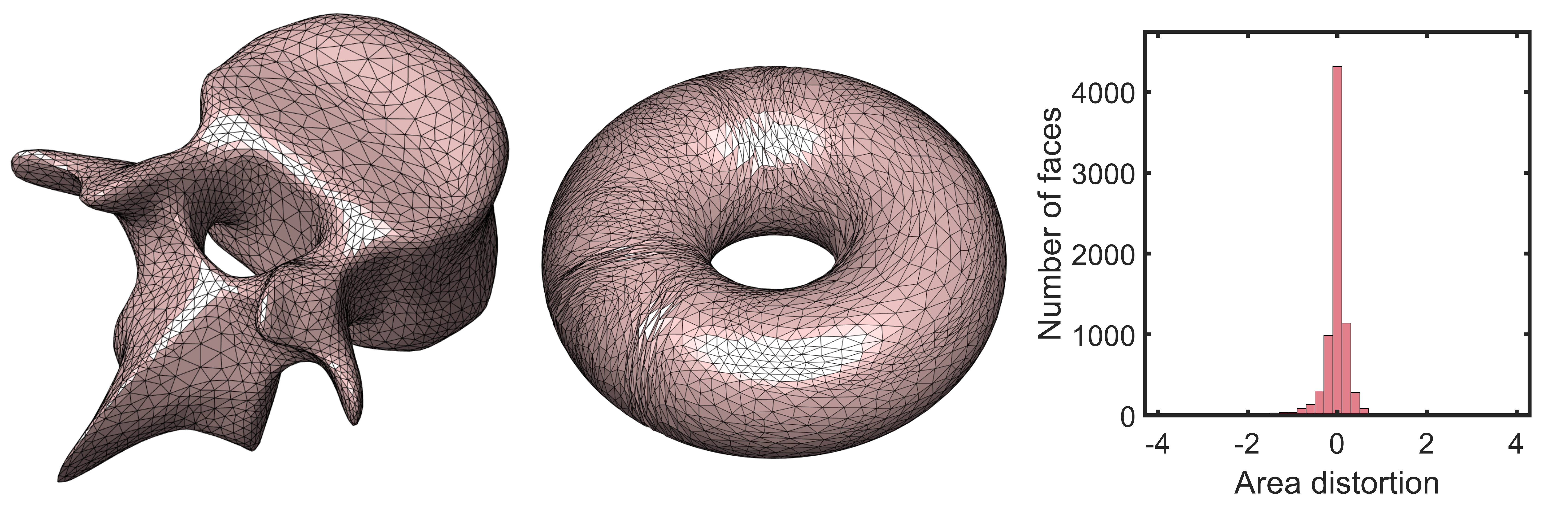}
    \caption{Toroidal area-preserving parameterization of the Vertebra model. Left to right: The input surface, the parameterization result by our algorithm, and the area distortion histogram.}
    \label{fig:parameterization_vertebra}
\end{figure}

Fig.~\ref{fig:parameterization_bracelet} shows an example of computing the toroidal area-preserving parameterization of the genus-one Bracelet surface model from~\cite{zhou2016thingi10k}. It can be observed that the original surface geometry is highly complex, with multiple sharp spatial variations in the local area and the surface curvature. Using our proposed method, we can easily parameterize the surface onto a standardized toroidal domain. To assess the area-preserving property of the parameterization, we consider the area distortion $d_{\text{area}}$ defined as
\begin{equation}
    d_{\text{area}}(f)(T) = \displaystyle \log \left(\frac{ (\text{Area}(T))/A}{(\text{Area}(f(T)))/A_f}\right),
\end{equation}
where $A = \sum_{T' \in \mathcal{F}}\text{Area}(T')$ and $A_f = \sum_{T' \in \mathcal{F}}\text{Area}(f(T'))$. In other words, we study the ratio between the original triangle face area and the triangle face area in the parameterization result after normalizing both the total surface area of the original surface and that of the final torus to be 1. It is easy to see that $d_{\text{area}}(f)(T) = 0$ if and only if $(\text{Area}(T))/A = (\text{Area}(f(T)))/A_f$, i.e., $f$ preserves the normalized face area of the triangle $T$. As the area distortion histogram in Fig.~\ref{fig:parameterization_bracelet} is highly central at 0, we can see that the toroidal parameterization obtained using our proposed method is highly area-preserving. Fig.~\ref{fig:parameterization_2_1} shows another genus-one surface model from~\cite{edelsbrunner2001180}, in which the surface is formed by wrapping an elongated tube and bending it significantly. Using our proposed method, we can easily unwrap the surface and obtain the toroidal area-preserving parameterization. Again, it can be observed from the area distortion histogram that the parameterization is highly area-preserving. In Fig.~\ref{fig:parameterization_rocker_arm}, we further consider parameterizing the genus-one Rocker Arm model. In this example, we can see that the triangle elements in the input surface are highly non-uniform. Nevertheless, using our proposed method, we can effectively obtain a highly area-preserving toroidal parameterization. In Fig.~\ref{fig:parameterization_vertebra}, we consider the genus-one Vertebra model. Even for such a model with multiple sharp features, it can be observed that our proposed method can produce a highly area-preserving toroidal parameterization.

In Table~\ref{tab:tdem_parameterization}, we summarize our experimental results and record the area distortion of the initial parameterization $h$ and the final toroidal density-equalizing parameterization $f$ obtained by our proposed method for various surface models. It can be observed that our proposed parameterization method significantly reduces the area distortion of the parameterization by over 85\% on average. Altogether, the experimental results show that our proposed toroidal parameterization method is applicable to a wide range of genus-one surfaces.

\begin{table}[t]
    \centering
\resizebox{\textwidth}{!}{
    \begin{tabular}{c|c|c|c|c}
       Example  & $\text{mean}(|d_{\text{area}}(h)|)$ & $\text{mean}(|d_{\text{area}}(f)|)$ & \% of improvement & Time (s)\\ \hline
       Bob (Fig.~\ref{fig:illustration}(b)) & 0.7006 & 0.0676 & 90.35\% & 6.3147 \\
       Bracelet (Fig.~\ref{fig:parameterization_bracelet}) & 0.4173 & 0.0974 & 76.67\% & 9.7246 \\
       Wrapped Tube (Fig.~\ref{fig:parameterization_2_1}) & 0.1865 & 0.0314 & 83.15\% & 18.2099\\
       Rocker Arm (Fig.~\ref{fig:parameterization_rocker_arm}) & 1.4689 & 0.1228 & 91.64\% & 7.5292 \\
       Vertebra (Fig.~\ref{fig:parameterization_vertebra}) & 1.5654 & 0.1601 & 89.77\% & 8.1411 \\
    \end{tabular}
    }
    \caption{The performance of our proposed toroidal density-equalizing parameterization method. For each example, we set the input population as the triangle face area to achieve a toroidal area-preserving parameterization. For each of the initial parameterization $h$ and final parameterization result $f$, we record the mean of the absolute value of the area distortion $\text{mean}(|d_{\text{area}}|)$. We also calculate the percentage of improvement in the area distortion to assess the area-preserving performance.}
    \label{tab:tdem_parameterization}
\end{table}

\begin{figure}[t]
    \centering
    \includegraphics[width=\linewidth]{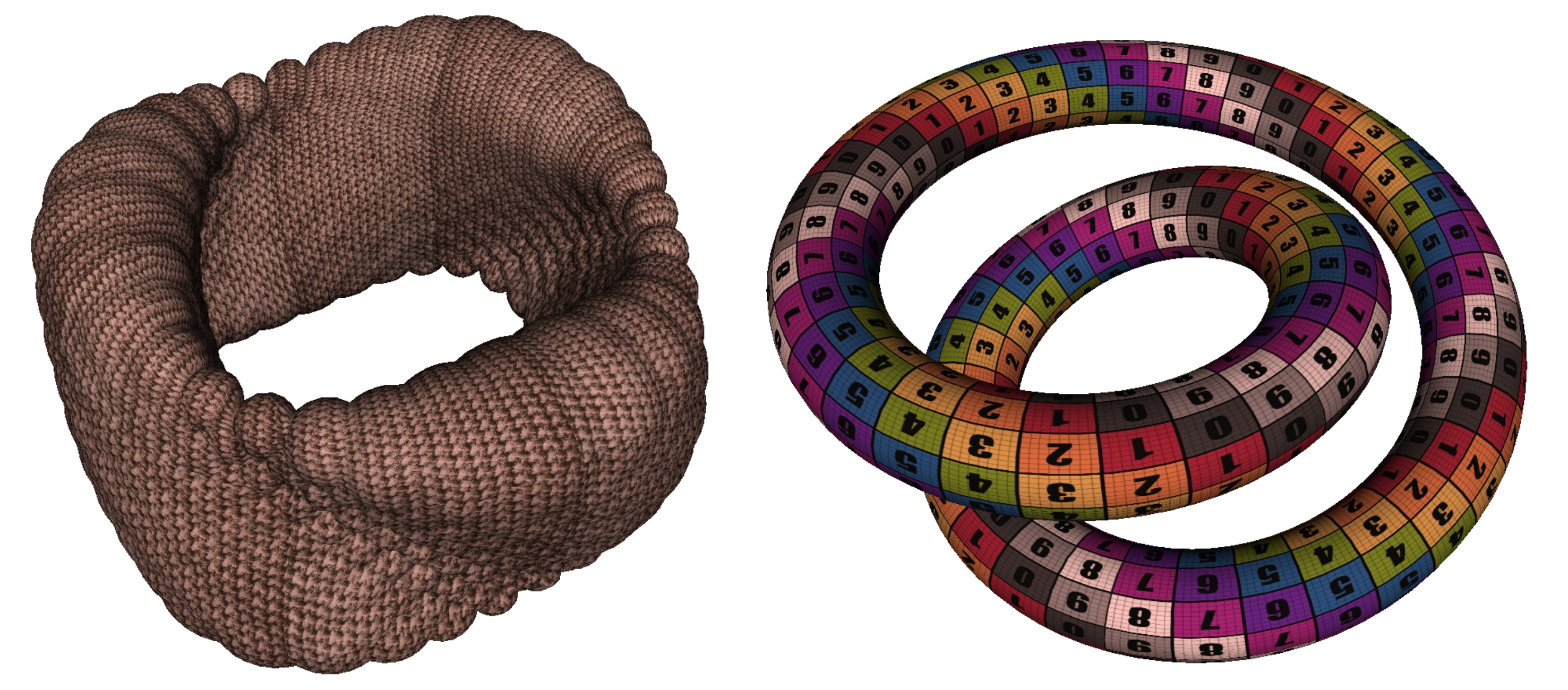}
    \caption{Texture mapping of genus-one surfaces using our proposed toroidal parameterization method. }
    \label{fig:texture}
\end{figure}

Using the proposed toroidal density-equalizing parameterization method, we can effectively perform seamless texture mapping for genus-one surfaces. Specifically, to design a texture on a given genus-one surface, we can map it onto a prescribed toroidal domain using the proposed parameterization method. Next, we can easily create texture patterns on the toroidal surface or its associated planar domain under the inverse toroidal projection $\phi^{-1}$. Finally, we can map the designed texture back to the given surface using the inverse of the toroidal parameterization. To illustrate this idea, in Fig.~\ref{fig:texture} we show genus-one surface texture mapping examples obtained using the above-mentioned approach. Because of the density-equalizing property of the proposed parameterization method, we can easily control the area change in different regions, thereby controlling the uniformity of the texture on the given surface. From the texture mapping examples, it can be observed that the textures are highly uniform. This shows that our method is well-suited for genus-one surface texture mapping.

\section{Conclusion and Discussion}\label{sec:conclusion}
In this paper, we have developed a novel method for computing density-equalizing maps for toroidal surfaces. We have shown that a large variety of toroidal shape deformation effects can be achieved by prescribing different population functions for the density diffusion process. We have also developed a toroidal density-equalizing parameterization method for general genus-one surfaces, which allows us to easily achieve toroidal parameterizations with controllable area changes. In particular, we have demonstrated the effectiveness of our method for computing toroidal area-preserving parameterizations.

As we have shown in our experiments, our proposed TDEM method can be flexibly applied to toroidal surfaces with different major and minor radii. Therefore, a natural next step is to extend the proposed toroidal parameterization method by incorporating an optimization procedure to further optimize the toroidal geometry and reduce not only the density-equalizing errors but also other geometric distortions of the parameterization results. Another possible future direction is to utilize the proposed methods for the shape analysis of genus-one surfaces.
  
\bibliographystyle{elsarticle-num}
\bibliography{reference}

\end{document}